%% file: main.tex
\documentclass[10pt, conference, letter]{IEEEtran}
\IEEEoverridecommandlockouts

\usepackage{amsmath, amsfonts, amssymb}
\usepackage{xcolor}
\usepackage{pifont}
\usepackage{comment}
\usepackage{graphicx} 
\usepackage{multirow} 
\usepackage{hhline} 
\usepackage{subcaption} 
\usepackage{adjustbox} 
\usepackage{tikz}
\usepackage{nicematrix}
\usepackage{tabularx}
\usepackage{adjustbox}
\usepackage{longtable}
\usepackage{booktabs}
\usepackage{caption}
\usepackage{threeparttable}

\usepackage{algorithm}
\usepackage{algpseudocode}
\usepackage{pbalance}
\usepackage{soul}

\definecolor{myblue}{HTML}{B8E0F2}
\definecolor{mygreen}{HTML}{C5E8D8}
\definecolor{mypeach}{HTML}{FFE6C7}

\usepackage{xcolor} \usepackage{pifont} \definecolor{darkgreen}{RGB}{0,190,0} \definecolor{softred}{RGB}{180,60,60}

\newcommand{\ymark}{\textcolor{darkgreen}{\ding{51}}}
\newcommand{\xmark}{\textcolor{softred}{\ding{55}}}

\algrenewcommand\alglinenumber[1]{\scriptsize\ttfamily\makebox[2em][r]{#1:}}

\algnewcommand\algorithmicforeach{\textbf{for each}}
\algdef{S}[FOR]{ForEach}[1]{\algorithmicforeach\ #1\ \algorithmicdo}

\title{
MIVE: A Minimalist Integer Vector Engine for Softmax LayerNorm and RMSNorm Acceleration
}
\author{
\IEEEauthorblockN{Kosmas Alexandridis and Giorgos Dimitrakopoulos}
\IEEEauthorblockA{
Integrated Circuits Lab, Electrical and Computer Engineering, Democritus University of Thrace (DUTH), Greece}}

\begin{document}

\maketitle

\begin{abstract}
The rapid growth of Large Language Models (LLMs) has intensified the need for specialized hardware accelerators that can satisfy stringent inference latency and power constraints. Although matrix multiplications dominate the overall computational workload, non-linear vector normalization operations, such as LayerNorm, RMSNorm and Softmax can become critical hardware bottlenecks. Existing accelerators typically implement these functions using dedicated hardware blocks, leading to duplicated resources and inefficient silicon utilization.
To address this limitation, we propose a Minimalist Integer Vector Engine  (MIVE), a programmable architecture capable of executing all three operations within a unified datapath. By exploiting common computational patterns across LayerNorm, RMSNorm and Softmax the proposed vector engine maximizes hardware sharing while reducing implementation overhead.
Physical ASIC implementation results show that MIVE provides comprehensive multi-function support while achieving higher area and hardware efficiency than most state-of-the-art standalone accelerators.
\end{abstract}

\section{Introduction}
The rapid growth of Large Language Models (LLMs) has driven the development of specialized hardware accelerators capable of meeting the stringent latency, throughput, and energy-efficiency requirements of modern inference workloads. These accelerators improve LLM execution through massively parallel systolic arrays, optimized dataflows, reduced-precision arithmetic, sparsity-aware computation, and hierarchical memory systems.

While such techniques effectively accelerate matrix operations, LLM performance also depends heavily on non-linear vector functions. In particular, Softmax, Layer Normalization (LayerNorm), and Root Mean Square Normalization (RMSNorm) are fundamental building blocks that play a critical role in maintaining numerical stability and preserving model quality. However, their reliance on costly element-wise transformations and vector-wide reductions, such as exponentiation, squaring, accumulation, square-root evaluation, and division, often makes them performance and efficiency bottlenecks in hardware implementations.

Numerous hardware accelerators have been proposed to reduce the cost of these operations by approximating or eliminating expensive exponential, division, and squaring computations. Division is commonly replaced by PWL reciprocal approximations~\cite{recip_approx}, power-of-two denominator quantization that enables shift-based scaling~\cite{softermax}, or logarithmic-domain formulations that transform division into subtraction~\cite{log-sum-exp}. Exponentials are similarly approximated through mathematical reformulations~\cite{softmex}, as well as polynomial~\cite{ibert}, fractional~\cite{softmax4gelu}, and PWL approximations~\cite{nn-lut} over reduced input domains. Finally, squaring overhead is reduced by operating on lower-bit-width representations of already quantized inputs~\cite{sole} or by employing specialized quantization schemes that enable efficient integer hardware~\cite{shao}.

Although these techniques substantially reduce hardware complexity, they typically target a single function. As a result, Softmax, LayerNorm, and RMSNorm are often implemented using separate accelerator blocks, leading to duplicated resources and inefficient silicon utilization.

To address this limitation we propose a novel Minimalist Integer Vector Engine that can execute LayerNorm, RMSNorm and Softmax within a single programmable architecture by exploiting their common computational primitives. This enables a hardware efficient datapath that can be re-used accross all three operations, eliminating the need for multiple dedicated accelerators.
The main contributions of this work are summarized as follows:

\begin{itemize}

\item LayerNorm, RMSNorm, and Softmax are decomposed into a common set of primitive operations, enabling their execution on low-cost shared hardware units.

\item A novel Minimalist Integer Vector Engine (MIVE) architecture is proposed, which evaluates Softmax, LayerNorm, and RMSNorm by mapping their primitive operations onto a unified and highly optimized datapath, maximizing hardware reuse.

\item Physical ASIC implementation results show that the proposed MIVE provides comprehensive multi-function support while achieving higher area and hardware efficiency than most state-of-the-art standalone accelerators.

\end{itemize}

\input{alg_restructure}
\input{hardware}
\input{evaluation}

\section{Conclusions}
This work presented a Minimalist Integer Vector Engine (MIVE) that supports Softmax, LayerNorm, and RMSNorm within a unified hardware architecture. By decomposing these normalization kernels into a common set of primitive operations, the proposed design enables extensive resource sharing through a programmable datapath. ASIC implementation results show that the proposed MIVE reduces hardware cost while enabling the execution of all three normalization functions on a shared architecture.

\bibliographystyle{IEEEtran}
\bibliography{refs}
\end{document}

%% file: alg_restructure.tex
\section{Normalization in LLM Inference}
\label{s:norm-llm}

LayerNorm, RMSNorm and Softmax are central to modern LLM inference, supporting activation stability, numerical robustness and attention computation. 

\subsection{Normalization Functions}
\input{algos}
For an input vector $\mathbf{x}$ of length $N$, LayerNorm computes
\begin{equation}
\text{LayerNorm}(\mathbf{x})_i =
\frac{x_i-\mu}{\sqrt{\epsilon+\sigma^2}}\gamma_i+\beta_i
\label{e:layernorm}
\end{equation}
where $\gamma_i$ and $\beta_i$ are learned parameters, while $\mu$ and $\sigma^2$ denote the mean and variance of the input vector, respectively:
\begin{equation}
\mu = \frac{1}{N} \sum_{j=1}^{N} x_j \qquad \sigma^2 = \frac{1}{N} \sum_{j=1}^{N} (x_j - \mu)^2
\end{equation}

RMSNorm removes mean subtraction and bias addition, reducing the computation to
\begin{equation}
\text{RMSNorm}(\mathbf{x})_i =
\frac{x_i}{\sqrt{\frac{1}{N}\sum_{j=1}^{N}x_j^2+\epsilon}}\gamma_i
\label{e:rmsnorm}
\end{equation}

Softmax normalizes exponentiated inputs into a probability distribution and is evaluated in its numerically stable form:
\begin{equation}
\text{Softmax}(\mathbf{x})_i =
\frac{e^{x_i-\max}}{\sum_{j=1}^{N}e^{x_j-\max}}
\label{e:stable-sm}
\end{equation}
where subtracting $\max$ bounds all exponential inputs to non-positive values. For clarity, $\epsilon$ in~\eqref{e:layernorm} and~\eqref{e:rmsnorm} is omitted in the following discussion.

\begin{algorithm}[t]
\caption{LayerNorm Correction (LNC)}
\label{alg:ln_corr_op}
\hspace*{\algorithmicindent} \textbf{Input:} $S_\text{new}$, $S_\text{old}$, $M_\text{new}$, $M_\text{old}$, $i$ \\
\hspace*{\algorithmicindent} \textbf{Output:} Corrected square sum $S_\text{old}$ and mean $M_\text{old}$
\begin{algorithmic}[1]
\State $S_\text{old}\gets S_\text{old}+S_\text{new}$ 
\State $S_\text{new} \gets a'''_k\cdot i+b'''_k$ \Comment{PWL $(i-1)/i$}
\State $M_\text{old} \gets M_\text{old} - M_\text{new}$ \Comment{$\Delta\mu$} 
\State $S_\text{new} \gets M_\text{old}\cdot S_\text{new}$ \Comment{$\Delta\mu\cdot(i-1)/i$}
\State $M_\text{new} \gets M_\text{new} + S_\text{new}$  \Comment{$M_\text{new}\gets\mu_i$} 
\State $M_\text{old} \gets {M_\text{old}}^2$
\State $S_\text{new} \gets a'''_k\cdot i+b'''_k$ \Comment{PWL $(i-1)/i$}
\State $M_\text{old} \gets S_\text{new}\cdot L$ \Comment{ $L\cdot (i-1)/i$ }
\State $S_\text{old}\gets M_\text{old}+S_\text{old}$
\State $M_\text{old}\gets M_\text{new}$
\end{algorithmic}
\end{algorithm}

\subsection{Iterative Computation of Normalization Functions}

To compute these normalization functions, hardware accelerators partition the input vector into sub-vectors of length $L$ and process them iteratively. Since the normalized output depends on global quantities over the entire vector $\mathbf{x}$, all partial results must be accumulated before the final normalization step. RMSNorm only requires the sum of squared elements and can therefore reduce each sub-vector independently. In contrast, LayerNorm and Softmax depend on a running mean or maximum, respectively, so previously accumulated values must be corrected whenever this reference changes.

For Softmax, the exponential sum is updated as:
\begin{align}
\label{e:corr_sm}
\text{Sum}_i =
\text{Sum}_{i-1}\cdot e^{\max_{i-1}-\max_i}+\sum_j e^{x_{ij}-\max_i}
\end{align}
where the previous partial sum is rescaled from the old maximum $\max_{i-1}$ to the updated maximum $\max_i$ before adding the contribution of the $i$-th sub-vector~\cite{softermax}.

\begin{algorithm}[t]
\caption{Softmax Correction (SMC)}
\label{alg:sm_corr_op}
\hspace*{\algorithmicindent} \textbf{Input:} $S_\text{new}$, $S_\text{old}$, $M_\text{new}$, $M_\text{old}$ \\
\hspace*{\algorithmicindent} \textbf{Output:} Corrected exponent sum $S_\text{old}$
\begin{algorithmic}[1]
\State $M_\text{old} \gets M_\text{old} - M_\text{new}$
\State $M_\text{old} \gets a_j{M_\text{old}}+b_j$ \Comment{PWL $e^x$}
\State $S_\text{old}\gets S_\text{old}\cdot M_\text{old}+S_\text{new}$
\end{algorithmic}
\end{algorithm}

For LayerNorm, the accumulated sum of squared deviations is corrected using the parallel variance update~\cite{par_var}. Assuming that $\text{Sum}_{i-1}$ and $\mu_{i-1}$ correspond to the first $i-1$ sub-vectors, while $\mu_{x_i}$ is the mean of the current sub-vector, the update is
\begin{align}
\label{e:corr_ln}
\text{Sum}_i &=\text{Sum}_{i-1}+\sum_j(x_{ij}-\mu_{x_i})^2+\frac{i-1}{i}L\cdot \Delta\mu^2 \\
\label{e:corr_mu}
\mu_i &= \mu_{x_i}+\frac{i-1}{i}\Delta\mu
\end{align}
where $\Delta\mu = \mu_{i-1}-\mu_{x_i}$ and the rightmost addition term accounts for the displacement between the previous running mean and the mean of the current sub-vector.

\subsection{Mapping Normalization to Primitive Operations}

Although LayerNorm, RMSNorm, and Softmax differ mathematically, their computations can be decomposed into primitive operations that map to the same hardware resources. These primitives can be grouped into two types.

The first type comprises element-wise operations used for input shifting, learned-parameter scaling, iterative correction, and function approximation. These computations include addition, subtraction, squaring, and general multiply-add operations. All of them can be computed by a single multiply-add operator by appropriately configuring its input operands, denoted as {\tt muladd}.

The second type comprises vector-wide reductions required during normalization. These include maximum and mean extraction, as well as the accumulation of squared or non-linear terms used to form the normalization factor. All such reductions are implemented by a single {\tt vecsum} operator.

Fig.~\ref{f:three-algs} shows the computation of LayerNorm, RMSNorm and Softmax using primitive operations. Any multiplication, addition or subtraction operation can be performed by the {\tt muladd} operator while mean, max and vector\_sum functions can be handled by the {\tt vecsum} operator. For LayerNorm and Softmax the corresponding correction routines are defined using primitive operation in Alg.~\ref{alg:ln_corr_op} and Alg.~\ref{alg:sm_corr_op} respectively.
The operations of these routines are handled by the {\tt muladd} operator since they contain only scalar affine operations, i.e. no vector reduction is needed.

%% file: algos.tex
\begin{figure*}[!t]
\centering
\scriptsize

\begin{minipage}[t]{0.32\textwidth}
\centering
\textbf{\textit{LayerNorm}}
\setlength{\abovedisplayskip}{2pt}
\setlength{\belowdisplayskip}{2pt}
\setlength{\jot}{1pt}
\vspace{0pt}
\begin{algorithmic}[1]
\For {$i = 1:\lceil \frac{N}{L} \rceil$}
\State $X\gets \textit{pop}\ X_i$
\State $M_\text{new}\gets \text{mean}(X)$
\State $\textit{push}\ X$ \Comment{Store $X_i$ back}
\State $X \gets X - M_\text{new}$
\State $X \gets {X}^2$
\State $S_\text{new}\gets \text{vector\_sum}(X)$ \Comment{$\sum_i (x_i-\mu)^2$}
\State $S_\text{old},M_\text{old}\gets \text{LNC}$ \Comment{Alg.~\ref{alg:ln_corr_op}}
\EndFor
\State $S_\text{old}\gets a'_kS+b'_k$ \Comment{PWL $1/\sqrt{S}$}
\State $S_\text{old}\gets S\cdot N^{-1}$
\For {$i = 1: \lceil \frac{N}{L} \rceil$}
\State $X\gets \textit{pop}\ X_i$
\State $X\gets X_i-M_\text{old}$ \Comment{$M_\text{old}$ is the global mean}
\State $X\gets X \cdot S$ \Comment{Normalize}
\State $X \gets \gamma_iX+\beta_i$
\Comment{LayerNorm($\mathbf x$)$_i$}
\State $\textit{push}\ X$
\EndFor
\end{algorithmic}
\end{minipage}
\hfill
\vrule
\hfill
\begin{minipage}[t]{0.32\textwidth}
\centering
\textbf{\textit{RMSNorm}}
\begin{algorithmic}[1]
\For {$i = 1: \lceil \frac{N}{L} \rceil$}
\State $X\gets \textit{pop}\ X_i$
\State $\textit{push}\ X$ \Comment{Store $X_i$ back}
\State $X \gets {X}^2$
\State $S_\text{new}\gets \text{vector\_sum}(X)$ \Comment{$\sum_i x_i^2$}
\State $S\gets S+S_\text{new}$
\EndFor
\State $S_\text{old}\gets a'_kS+b'_k$ \Comment{PWL $1/\sqrt{S}$}
\State $S_\text{old}\gets S_\text{old}\cdot N^{-1}$
\For {$i = 1: \lceil \frac{N}{L} \rceil$}
\State $X\gets \textit{pop}\ X_i$
\State $X\gets X \cdot S$ \Comment{Normalize}
\State $X\gets \gamma_iX$ \Comment{RMSNorm($\mathbf x$)$_i$}
\State $\textit{push}\ X$
\EndFor
\end{algorithmic}
\end{minipage}
\hfill
\vrule
\hfill
\begin{minipage}[t]{0.32\textwidth}
\setlength{\abovedisplayskip}{2pt}
\setlength{\belowdisplayskip}{2pt}
\setlength{\jot}{1pt}
\textbf{\textit{Softmax}}
\centering
\begin{algorithmic}[1]
\For {$i = 1:\lceil \frac{N}{L} \rceil$}
\State $X\gets \textit{pop}\ X_i$
\State $M_\text{new}\gets \text{max}(X, M_\text{old})$

\State $\textit{push}\ X$ \Comment{Store $X_i$ back}
\State $X \gets X - M_\text{new}$
\State $X \gets a_kX+b_k$ \Comment{PWL $e^{x}$}
\State $S_\text{new}\gets \text{vector\_sum}(X)$ \Comment{$\sum_i e^{x_i-\max}$}
\State $S_\text{old}\gets\text{SMC}$ \Comment{Alg.~\eqref{alg:sm_corr_op}}
\State $M_\text{old}\gets M_\text{new}$
\EndFor
\State $S_\text{old}\gets a''_kS+b''_k$ \Comment{PWL $1/S$}
\For {$i = 1: \lceil \frac{N}{L} \rceil$}
\State $X\gets \textit{pop}\ X_i$
\State $X\gets X - M_\text{old}$ \Comment{$M_\text{old}$ is the global max}
\State $X \gets a'_kX+b'_k$ \Comment{PWL $e^{X}$}
\State $X\gets X \cdot S$ \Comment{Softmax($\mathbf x$)$_i$}
\State $\textit{push}\ X$
\EndFor
\end{algorithmic}
\end{minipage}
\caption{LayerNorm, RMSNorm, and Softmax computation, expressed with primitive operations.}
\label{f:three-algs}
\end{figure*}

%% file: hardware.tex
\section{Minimalist Integer Vector Engine}
\label{s:hardware}

This section introduces the proposed Minimalist Integer Vector Engine (MIVE), a programmable architecture that executes LayerNorm, RMSNorm, and Softmax using shared hardware resources.
MIVE is designed around the multiply-add operator {\tt muladd}, and the vector reduction operator {\tt vecsum}, identified in the previous section. Rather than instantiating separate datapaths for LayerNorm, RMSNorm, and Softmax, MIVE exposes a programmable execution flow in which the primitive operations shown in Fig.~\ref{f:three-algs}
are computed using shared {\tt muladd} and {\tt vecsum} hardware units.

\begin{figure}[!h]
    \centering
    \includegraphics[width=0.98\columnwidth]{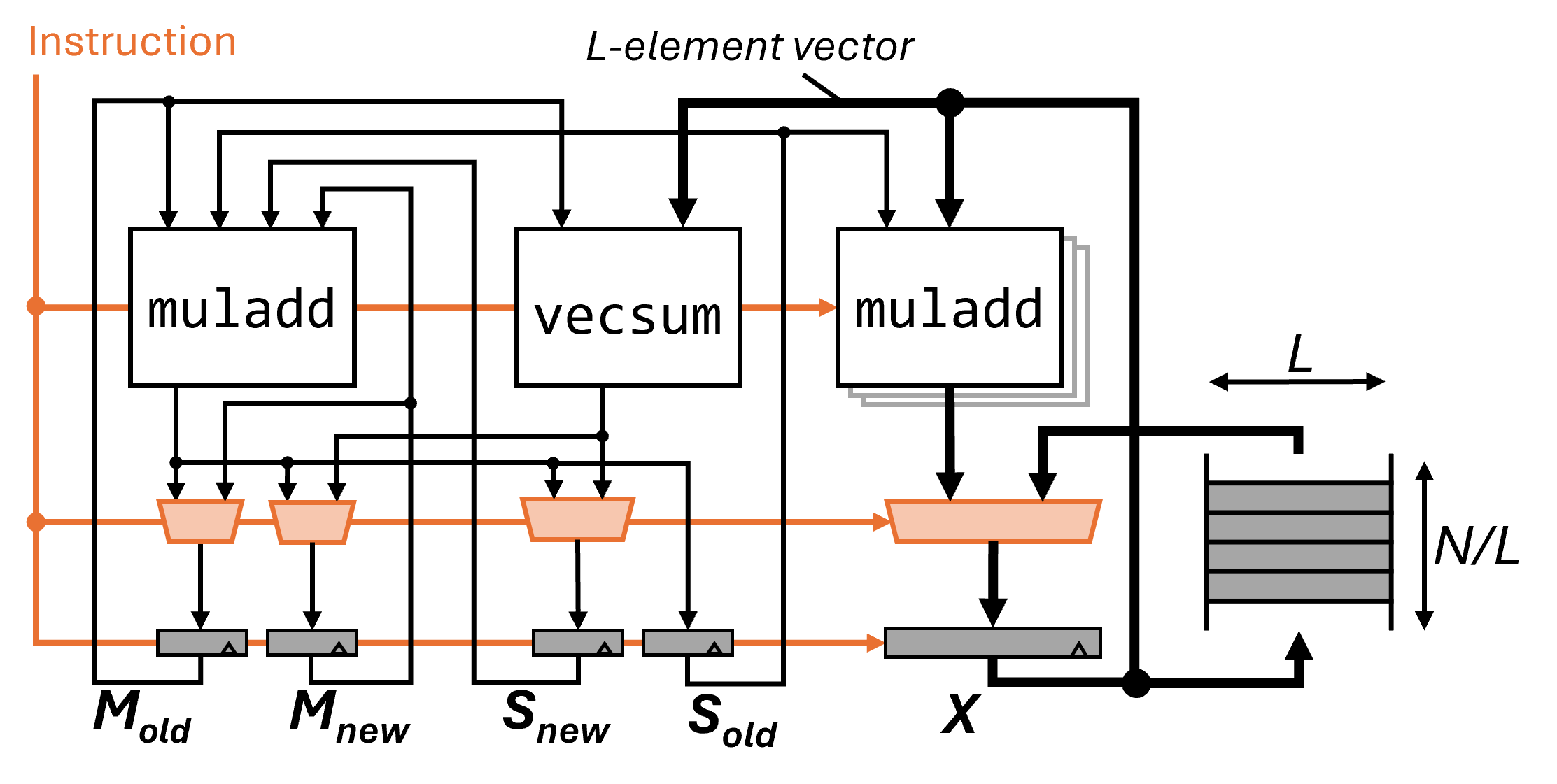}
    \caption{Hardware organization of the proposed MIVE. }
    \label{f:hw}
\end{figure}

Motivated by modern LLM accelerator paradigms~\cite{sole}, the proposed Minimalist Integer Vector Engine targets INT8 quantized LLM inference. All operations use integer arithmetic, with fixed-point representations employed where needed. To enable PWL approximations on an integer multiply-add datapath, an architecture similar to~\cite{gqa} is employed. Intermediate values are stored in sufficiently wide integer formats to avoid overflow, while outputs retain the same numerical format as the inputs.

The proposed Minimalist Integer Vector Engine (MIVE) is controlled through an ISA-level interface, where instructions encode both the target primitive and the operation to be executed. The instruction bits are used directly to drive the select signals of the arithmetic units and to control data movement across the datapath which avoids redundant instruction decoding logic.

Vector element-wise operations are computed by an array of {\tt muladd} units operating in parallel across the vector lanes. Each unit contains a shared multiplier-adder datapath, while both addition and subtraction are supported by conditionally complementing the right-hand-side operand. The PWL coefficients used for exponent ($e^{x_i-\max}$) approximations are stored in local ROMs within each {\tt muladd} unit.
A local vector register $X$ feeds the array and stores its outputs, enabling consecutive operations on the same vector without accessing slower on-chip buffers. New sub-vectors are loaded into $X$, while final normalized outputs are written back to the on-chip buffers before being transferred to external memory.

A single scalar {\tt muladd} unit decouples scalar operations from vector element-wise execution. It uses the same multiply-add datapath as the vector {\tt muladd} units and stores PWL coefficients for exponent approximation, $1/\Sigma$, $1/\sqrt{\Sigma}$, and the LayerNorm correction factor $(1-j)/j$ in Alg.~\ref{alg:ln_corr_op}. This unit is mainly used for correction routines and normalization-factor computation.
It interfaces with four local registers, $M_\text{old}$, $M_\text{new}$, $S_\text{old}$, and $S_\text{new}$, which store mean/maximum values, running sums, and intermediate correction results. These registers are local to MIVE and do not directly access external memory.

Finally, vector-wide reductions are performed by {\tt vecsum}, a binary-tree unit whose nodes support both addition and subtraction. The subtraction result enables pairwise maximum selection, allowing the same tree to perform accumulation and maximum reduction. The unit reads operands from $X$ and writes the scalar result to either $M_\text{new}$ or $S_\text{new}$.

%% file: evaluation.tex
\section{Evaluation}

Experimental results highlight the hardware savings of the proposed Minimalist Integer Vector Engine and evaluate its impact on real machine-learning applications.

\subsection{Hardware complexity evaluation}
\begin{figure}[!h]
    \centering
    \includegraphics[width=0.8\columnwidth]{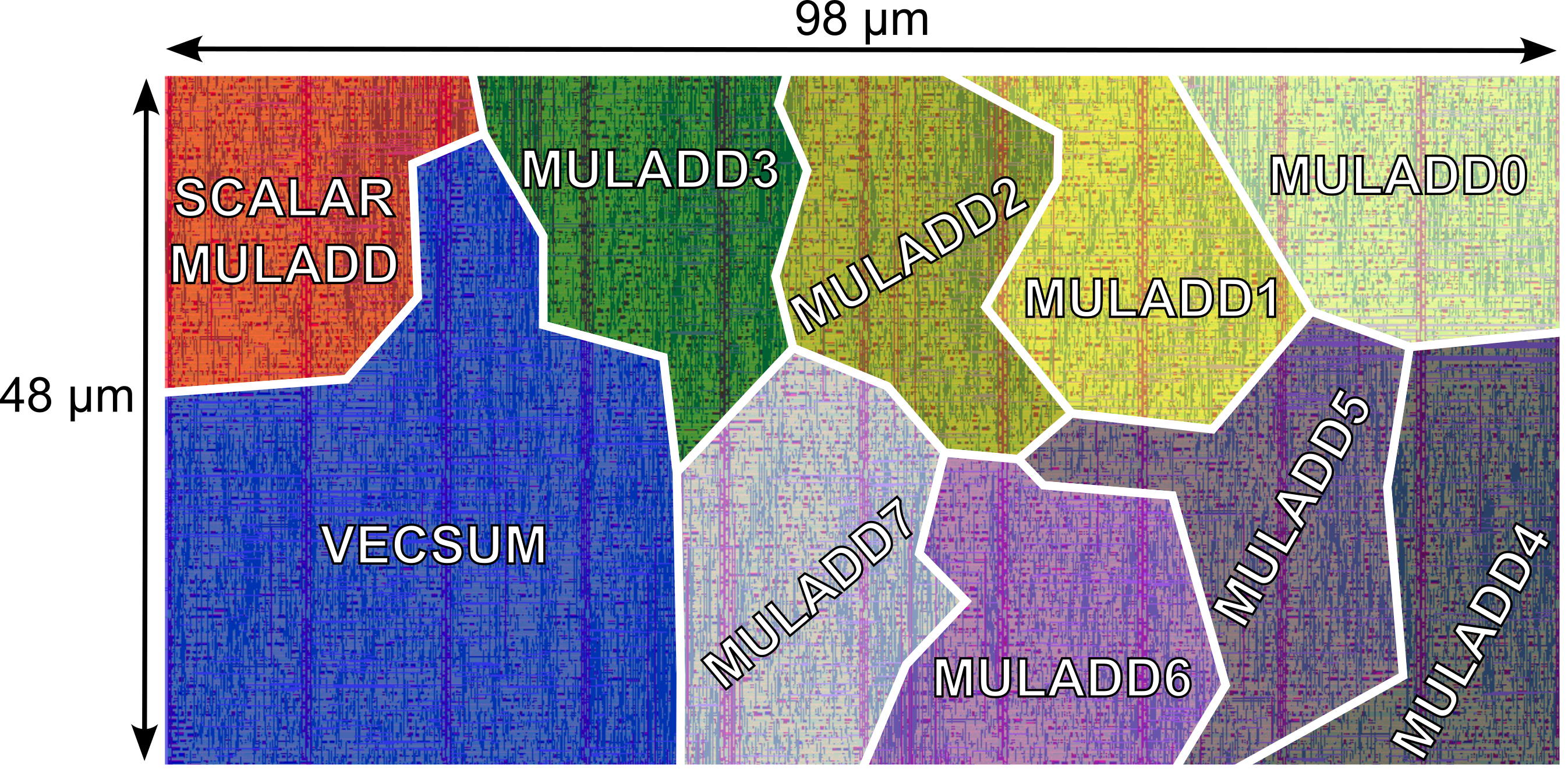}
    \caption{The physical layout of the proposed Minimalist Integer Vector Engine for hardware length of 8 parallel elements.}
    \label{f:phy}
\end{figure}

To quantify the hardware savings, we implemented the proposed MIVE design in SystemVerilog RTL.
Physical synthesis and Power estimation was carried out using Cadence's digital implementation tools using a 28-nm standard-cell technology and targeting a clock frequency of 2GHz. where the physical design is show in Fig.~\ref{f:phy}. The reported power reflects average consumption measured during inference on various benchmarks using the {\tt lm-evaluation-harness} framework~\cite{lm-eval}.

\begin{table*}[!ht]
    \caption{Comparison of the proposed MIVE with SoTA designs for hardware length of $L$ = 8 parallel elements.}
    \label{t:sota_comp}
    \centering
    \begin{threeparttable}     
    \centering
    \begin{tabular}{|c||c|c|c|c|c|c|}
    \cline{2-7}
        \multicolumn{1}{c|}{} & Spagnolo et al. &  Kim et al. & Softm$e^x$& Shao et al. & SOLE & Proposed \\
        \multicolumn{1}{c|}{} & \cite{spagnolo} &  \cite{kim} &~\cite{softmex} &~\cite{shao} &~\cite{sole} & MIVE \\ \hline
        Process (nm) & 28 & 28 & 45 & 28 & 28 & 28 \\ \hline
        Precision & INT-16 & INT-16 & INT-16 & INT-12 & INT-8 & INT-8 \\ \hline
        Frequency (GHz) & 2.50 & 2.50 & 1.70 & 1 & 1.50 & 2 \\ \hline
        Area ($\mu\text{m}^2$) & 3595 & 24900 & 13980 & 29472 & 7622 & 4811 \\ \hline
        Power (mW) & 2.40 & 52.46 & 58.94 & 16.64 & 18.80 & 9.50 \\ \hline
        Throughput (GOPS) \tnote{a} & 25 & 20 & 13.60 & 8 & 12 & 16 \\ \hline
        Area Eff. (GOPS$/\text{mm}^2$) & 695.4 & 80.3 & 97.3 & 27.1 & 157.4 & 332.6 \\ \hline
        Power Eff. (GOPS$/$mW) & 10.4 & 0.4 & 0.2 & 0.5 & 0.6 & 1.7 \\ \hline
        \hline
        LayerNorm & \xmark & \xmark & \xmark &\ymark & \ymark & \ymark \\ \hline
        RMSNorm & \xmark & \xmark & \xmark &\xmark & \xmark & \ymark \\ \hline
        Softmax & \ymark & \ymark & \ymark &\xmark & \ymark & \ymark \\ \hline
    \end{tabular}
    \begin{tablenotes}
    \footnotesize
    \item[a] Computed as (hardware length $\times$ operating frequency).
    \end{tablenotes}
    \end{threeparttable}
\end{table*}

Table~\ref{t:sota_comp} compares the proposed MIVE with SoTA designs for dedicated normalization acceleration. MIVE provides high hardware efficiency while supporting Softmax, LayerNorm, and RMSNorm on a shared datapath. It achieves $332.6$ GOPS$/\text{mm}^2$, improving area efficiency by $2.1\times$ over~\cite{sole} and by $4.1\times$, $3.4\times$, and $12.3\times$ over~\cite{kim}, ~\cite{softmex}, and~\cite{shao}, respectively. In terms of power efficiency, MIVE reaches $1.7$ GOPS$/\text{mW}$, corresponding to a $2.8\times$ improvement over~\cite{sole} and $3.4\times$ over~\cite{shao}. Although~\cite{spagnolo} reports higher hardware efficiency, in both area and power it only supports Softmax, while the proposed MIVE targets all three normalization functions.

\subsection{Evaluation of LLM accuracy}

To assess the impact of the PWL approximations employed by MIVE, we run inference for two LLM workloads shown in Table~\ref{t:benchmarks}. For each model we ran two variants: An accurate Python model that uses FP16 precision throughout its execution and a pure INT8 variant quantized using the SMOOTHQUANT scheme~\cite{smq} that uses the proposed MIVE engine to compute the respective normalization operations. OPT-30B~\cite{opt} uses a combination of Softmax and LayerNorm while Llama2-7B~\cite{llama} uses RMSNorm instead.
For OPT-30B on LAMBADA~\cite{lambada}, the INT8+MIVE configuration reduces accuracy only marginally, from $81\%$ to $80\%$. Similarly, for Llama2-7B on Wikitext~\cite{wikitext}, perplexity increases slightly from $5.8$ to $6.0$. These results indicate that the proposed approximations introduce only limited accuracy degradation and do not significantly affect model-level performance.

\begin{table}[!h]
\caption{Performance of LLM models}
    \label{t:benchmarks}
    \centering
    \begin{adjustbox}{width=1\columnwidth}
    \begin{tabular}{|c|c|c|c|}
    \hline
        \multicolumn{2}{|c|}{OPT-30B/LAMBADA} &  \multicolumn{2}{c|}{Llama2-7B/Wikitext} \\
        \multicolumn{2}{|c|}{accuracy(\%) $\uparrow$} &  \multicolumn{2}{c|}{perplexity $\downarrow$} \\ \hline
        FP16 & INT8+VNU  & FP16 & INT8+VNU \\ \hline
        81 & 80 & 5.8 & 6.0 \\ \hline
    \end{tabular}
    \end{adjustbox}
\end{table}